\magnification=1000


\hsize=190truemm \advance\hsize by-33truemm
\vsize=280truemm \advance\vsize by-50truemm

\parindent=0pt
\hfuzz=50pt
\parskip=5pt
\input eplain.tex
\leftdisplays
\input graphicx
\input opmac

\def\Myblue{\setcmykcolor{1.0 0.25 0. 0.}}

\vskip -0.5truecm
\includegraphics[width=3.0cm, angle=0]{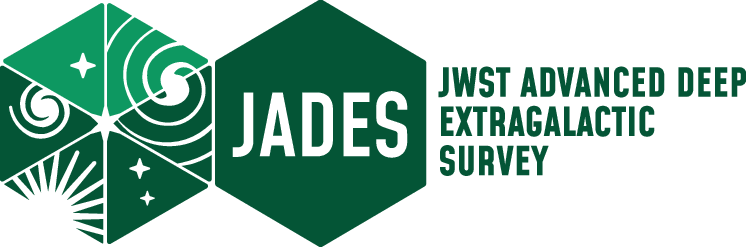}

\bigskip
\centerline{\bf On-orbit Degradation of the JWST/NIRSpec Micro-Shutter Array:} 
\centerline{\bf Multiplexing and Target Acquisition} 
\medskip
\centerline{Peter Jakobsen, Cosmic Dawn Center, Niels Bohr Institute, University of Copenhagen}
\centerline {Issue 1.01, 7 May 2024}
\bigskip

\centerline{\vbox{\hsize=15truecm 
{\bf Abstract:} A statistical analysis of the impact of the diminishing number of operational shutters experienced by the JWST/NIRSpec Micro Shutter Array since Commissioning is presented. It is shown that the number of high priority science targets that NIRSpec is able to observe simultaneously has so far decreased by 3.1\%. Of greater concern, however, is NIRSpec's diminished ability to carry out autonomous MSATA target acquisition, which is more sensitive to the loss of shutters than is the multiplexing. In the flagship case of MSA observations of deep fields, the number of pointings at which it is not possible to reach the required minimum number of 5 Valid Reference Stars has increased from 4.9\% to 6.3\% and is beginning to become noticeable. Similarly, the number of higher risk target acquisitions that need to be carried out with fewer than the maximum allowed number of 8 Reference Stars has grown from 27\% to 31\%.}}

\medskip
{\bf 1. Introduction}

Prior to launch, 15.0\% of the  shutters in the unvignetted part of NIRSpec's Micro Shutter Array (MSA) were known to be non-operational for various reasons. Following launch and Commissioning, this fraction had risen to 16.7\%. During the latest tally performed in June 2023, the non-functional shutter fraction had increased further to 18.3\%, partly because a number of shutter rows and columns needed to be intentionally masked out in order to avoid electrical shorts appearing in the device. The impact of this degradation is usually thought of in terms of the resulting loss in multiplexing capability experienced by NIRSpec when operating in MSA mode. However, this is not the full story. The purpose of this note is to also quantify the ability of NIRSpec to successfully perform autonomous target acquisition in MSA mode, and demonstrate that this functionality is in fact more impacted by the loss of operational shutters than is the number of targets that can be observed simultaneously.

\midinsert 
\vskip 0.4truecm
\hbox{\hskip 3.3truecm {\vbox{\halign{
#\hfil&   \quad\quad #\hfil\cr 
\noalign{\hskip 0.0truecm {\bf Table 1:} Reference MSA Files}
\noalign{\medskip}
\noalign{\hrule}
\noalign{\medskip}
March 2022:& \cr
Checkerboard:& \verbatim NRS_MSOP_CHK_2.0.1_20220312T0000.msl|endverbatim\cr
Short Mask:&    \verbatim NRS_MSOP_SM_3.0.0_20220326T0000.msl|endverbatim\cr
Vignetting:&   \verbatim NRS_MSOP_VIG_2.0.0_20211225T0000.msl|endverbatim\cr
\noalign{\medskip}
June 2023:& \cr
Checkerboard:& \verbatim NRS_MSOP_CHK_2.0.1_20230115T0000.msl|endverbatim\cr
Short Mask:&    \verbatim NRS_MSOP_SM_4.0.0_20230606T0000.msl|endverbatim\cr
Vignetting:&   \verbatim NRS_MSOP_VIG_2.0.0_20211225T0000.msl|endverbatim\cr
\noalign{\medskip}
\noalign{\hrule} 
\noalign{\medskip}
}} } }
\endinsert

{\bf 2. Reference tables used in the analysis}

The MSA operability map tracking the state of each of the MSA's $4\times365\times171=249\,660$ shutters is constructed from three separate so-called MSOP fits tables resident in the project database [1]. The primary measurement is the so-called `checkerboard' table which is generated from a series of exposures carried out in imaging mode using the internal lamps, with the MSA commanded to several fine-grained patterned configurations from which the functional, failed closed or failed open state of each shutter can be inferred from its projection on the detector array. The checkerboard table is supplemented by the so-called `short mask' table listing the shutters situated in rows or columns that it has been necessary to make non-operational due to the presence of electrical and/or optical shorts appearing in the array quadrants. Finally the `vignetting' map specifies the outermost ($\simeq12$ shutter rows at the top and bottom and $\simeq7$ columns along the outer sides of the intentionally over-sized MSA) that are occulted by the $3.6\times3.4$~arcmin$^2$ physical field stop located in the OTE focal surface upstream of the MSA.

When constructing the overall MSA operability map from these three tables, it is important that the entries for each shutter be combined in the correct hierarchical order such that the short mask renders both functional and failed closed shutters in the checkerboard table non-functional permanently closed, except for the failed opens which remain so, and the vignetting map renders all occulted shutters non-functional permanently closed regardless of their actual functional state.

Since the primary concern is the MSA degradation that has occurred in orbit,  the MSA state measured during Commissioning in March 2022 is adopted as the baseline and compared to the most recent census performed in June 2023. The corresponding 
MSOP tables are listed in Table~1. Note that the vignetting reference file is the same on both dates since the vignetting appears to have been very stable since launch. At the time of writing, the most recent short mask available is the one introduced in June 2023 which was prompted by several shorts appearing in quadrants Q1 and Q2.

\midinsert 
\vskip 0.4truecm
\hbox{\hskip 2.5truecm {\vbox{\halign{
#\hfil&   \quad\quad \hfil#& \ \hfil#& \quad \hfil#& \ \hfil#& \quad \hfil#& \ \hfil#\cr 
\noalign{\hskip 0.0truecm {\bf Table 2:} MSA Shutter Operability Statistics}
\noalign{\medskip}
\noalign{\hrule}
\noalign{\medskip}
Shutter State& March 2022\phantom{i}& \%\phantom{i}& June 2023& \%\phantom{i}\cr
\noalign{\medskip}
\noalign{\hrule}
\noalign{\medskip}
Functional   & 188006&  75.3&  184391& 73.9\cr
Failed Closed&  14916&   6.0&   16561&  6.6\cr
Masked Out&     22694&   9.1&   24664&  9.9\cr
Vignetted    &  24024&   9.6&   24024&  9.6\cr
Failed Open  &     20&      &      20&     \cr
\noalign{\medskip}
Number of Viable Slitlets  &     111847&  &   108329&     \cr
Number of Viable TA Windows&  103293& &      98415&      &     \cr
\noalign{\medskip}
\noalign{\hrule} 
\noalign{\medskip}
}} } }
\vskip -0.1truecm
\endinsert

{\bf 3. Failed Shutter Statistics}

The statistics summarizing the operational state of the MSA on March 2022 and June 2023 extracted from the above reference files are listed in Table~2. For the purpose of this note we distinguish between functional, failed closed, masked out, vignetted and failed open shutters. The listed table columns therefore sum to $4\times365\times171$ and the percentages listed refer to the entire array.

It is apparent that the total number of functional shutters decreased by 3615  over the 16 month timespan between the two epochs. This 1.9\% degradation was caused in roughly equal measure by 1645 new isolated failed closed shutters appearing on the MSA, and needing to intentionally mask off 1970 previously operational shutters due to shorts. 

{\bf 4. Impact on NIRSpec Multiplexing}

As explained in detail in [2], the multiplexing capability of NIRSpec in MSA mode is to first order dictated by the probability that any candidate target located within the unvignetted area of the MSA falls on a shutter capable of serving as a fully functional three shutter tall slitlet that can be commanded opened and used to perform a standard nodded observation of the target. The second-to-last line in Table~1 provides the number of such Viable Slitlets contained on the MSA at the two epochs. Figure~1 displays the map of the Viable Slitlets for the June 2023 date. It is evident that the 1.9\% MSA degradation experienced between March 2022 and June 2023 in the number of operational shutters translates into a 3.1\% loss in the number of shutters capable of serving as the central shutters of Viable Slitlets. The latter is the relevant metric for gauging how many high priority targets NIRSpec can observe simultaneously, given that these get placed on the MSA first and therefore fall on the linear portion of the multiplexing curve. Note that this relative drop in multiplexing is larger than the relative loss of functional shutters since each new failed shutter can give rise to the loss of up to three Viable Slitlets. 

\midinsert
\vskip 0.0truecm
\centerline {\includegraphics[width=10.2cm, angle=0]{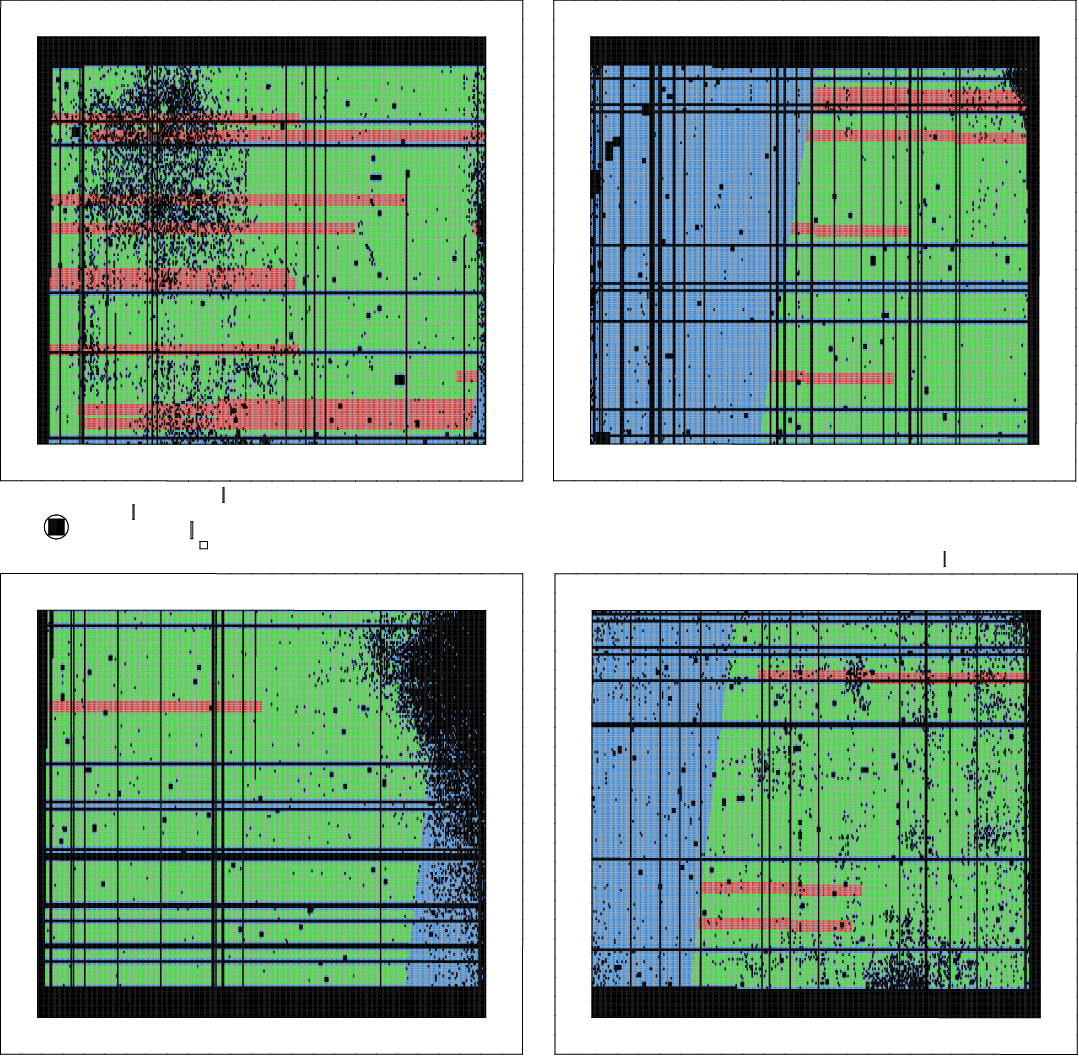}}
\vskip 0.5truecm
\centerline{\vbox{\hsize=13.5truecm {\bf Figure 1:} The MSA Viable Slitlet map in PRISM mode as of June 2023. Clockwise from upper left the quadrants are Q3, Q1, Q2 and Q4. The 65\,249 non-operational shutters that are either failed closed, masked out or vignetted by the NIRSpec field stop are depicted in black. The 108\,329 shutters capable of serving as the central shutters of fully functional three shutter tall slitlets are marked in {\localcolor \Green green}. These span 48\% of the un-vignetted area of the MSA. The 17\,340 shutters excluded due to their spectra clashing with those from failed open shutters are marked in {\localcolor \Red red}. Shutters depicted in {\localcolor \Myblue blue} are operational but excluded from use, either because they do not form the center of a fully functional 3-slitlet, or because their PRISM spectra project onto the detector gap.}}
\endinsert

\medskip
{\bf 5. Impact on NIRSpec Target Acquisition}

\midinsert
\vskip 0.0truecm
\centerline {\includegraphics[width=10.2cm, angle=0]{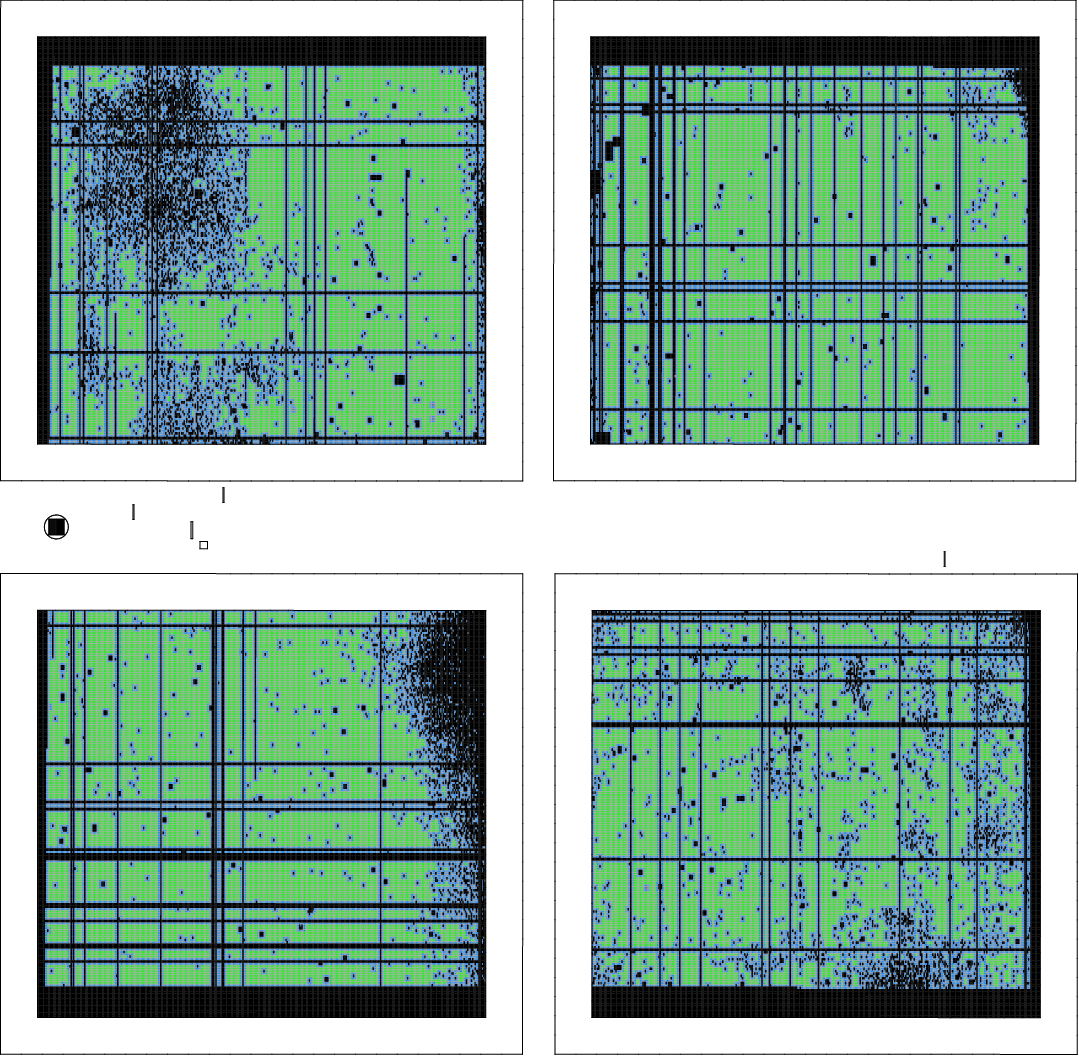}}
\vskip 0.5truecm
\centerline{\vbox{\hsize=13.5truecm {\bf Figure 2:} Status of the current MSA from the perspective of NIRspec target acquisition as of June 2023. Clockwise from upper left the quadrants are Q3, Q1, Q2 and Q4. Shutters that are non-operational due to their being failed closed, masked out or vignetted by the NIRSpec field stop are depicted in black, and the 20 permanently failed open shutters in {\localcolor \Red red}. The 98\,415 shutters plotted in {\localcolor \Green green} -- representing 43.6\% of the un-vignetted area of MSA -- are capable of serving as the centers of $5\times3$ fully operational shutter windows needed for acquiring Reference Stars during target acquisition. Shutters depicted in {\localcolor \Myblue blue} are operational, but do not form the centers of viable target acquisition windows.}}
\endinsert

The NIRSpec MSATA target acquisition (TA) process is described in [3]. Briefly summarized, the scheme relies on the user to identify as part of the observation planning a number of so-called Reference Stars within the MSA field of view that the onboard software can locate and centroid in dedicated TA exposures taken in NIRSpec's imaging mode, and use to calculate the corrective slew needed to adjust the telescope pointing in pitch, yaw and roll such that the MSA is oriented on the sky as intended, and the science  targets of interest are placed within their designated commanded open slitlets. Note that this process is separate from the selection of Guide Stars used by the Fine Guidance Sensor to hold the telescope pointing steady during the observation. 

To allow for a range in Reference Stars to be employed, the user may, depending on the particulars of the field being observed, choose to carry out the target acquisition in  one of three broadband TA filters and employ one of several fixed exposure times. The user embeds the list of reference star candidates into the target catalog along with their estimated TA filter magnitudes. Upon entering the catalog into the Astronomers Proposal Tool (APT), the system then selects a miniumum of 5 and a maximum of 8 valid Reference Stars to be included in the command load and used during the initial TA for the visit. The more candidate Reference Stars that are accepted by the APT, the more likely the TA is to succeed. In addition to selecting candidate Reference Stars whose listed magnitudes lie within the allowable brightness range for the TA filters, the APT also requires that each selected Reference Star at the nominal pointing falls within an operational shutter that forms the center of a fully operational 5 shutter wide and 3 shutter tall  window of fully operational shutters. This last requirement is needed to assure that the Reference Star can be located and centroided within its window in the acquisition image, and isn't obscured by a failed closed shutter.

The last row in Table~2 lists the number of such $5\times3$ Viable TA Windows available on the MSA at the two epochs. Figure~2 shows a map of the TA active area as of June 2023. It is evident that the number of Viable TA Windows decreased by 4.7\% between the two epochs. It should be noted that this loss in active TA area is amplified even more w.r.t. the 1.9\% loss in functional shutters compared to the Viable Slitlet count since each new failed shutter can result in the loss of up to 15 nearby target acquisition windows. For this reason NIRSpec target acquisition is more threatened by long term degradation of the MSA than is the multiplexing.  

It is instructive to explore the impact of the loss in target acquisition capability in further detail. As an example close to heart, we adopt the case of deep observations of the most remote galaxies in the GOODS-N and GOODS-S fields carried out during the large collaborative JADES NIRCam/NIRSpec GTO program [4]. This choice should be representative of other deep field programs such as CEERS, UNCOVER, etc. Such deep NIRSpec observations are particularly challenging since the fields involved contain far too few stars to use as Reference Stars, requiring instead that sufficiently `pointy' faint galaxies having brightnesses matching the broadest available (CLEAR) filter and the longest possible (NRSRAPIDD6) TA exposure time be used. At the start of JADES, the task of identifying suitable candidate galaxies on the basis of the HST data alone was not a trivial exercise given the finite spatial resolution and wavelength range of the available HST images (not to mention their relatively inaccurate astrometry). This selection process became considerably more robust once the GOODS fields had been carefully mosaicked through NIRCam imaging and properly referenced to the GAIA reference frame. To date, no  NIRSpec target acquisition carried out as part of JADES has failed. 

One key statistic for NIRSpec target acquisition is the average density of available candidate Reference Stars on the sky. The JADES program has established the density of galaxies suitable for target acquisition in GOODS-S to be  $\Sigma_{\scriptscriptstyle R\!S}=2.26\pm0.21$~arcmin$^{-2}$. The number of such candidate Reference Stars present on the MSA at any given pointing will therefore be Poisson distributed with parameter
$$\lambda=\Sigma_{\scriptscriptstyle R\!S} \Omega_{\scriptscriptstyle M\!S\!A} = 20.1 \eqno(1)$$
where $\Omega_{\scriptscriptstyle M\!S\!A}=8.89$~arcmin$^2$ is the solid angle extended by the non-vignetted portions of the MSA.

The probability that any one of these randomly placed candidate Reference Stars will land in a Viable TA Window is given by the relative area spanned by the central shutters of all such TA windows, which per Table~2 in June 2023 was
$$p_{\scriptscriptstyle T\!A}= 98\,415/(4\times365\times171-24\,024)=0.436 \eqno(2)$$
The number of Valid Reference Stars landing in Viable TA windows will therefore also be Poisson distributed, but with a `thinned' Poisson parameter
$$\tilde{\lambda}= p_{\scriptscriptstyle T\!A} \lambda = 8.76 \eqno(3)$$

The probability that there will be less that $m$ Valid Reference Stars available at any given pointing is therefore given by the cumulative Poisson distribution
$$p(k<m) =  \sum_{j=0}^{m-1} e^{-\tilde{\lambda}}{{\tilde{\lambda}^j}\over{j!}} = Q(m,\tilde{\lambda}) \eqno(4)$$
where $Q(n,x)$ is the incomplete Gamma function. 

The APT will reject any observation if there are less than $m=5$ Valid Reference Stars available at the pointing of the visit tied to the target acquisition. The probability of this occurring as a function of $p_{\scriptscriptstyle T\!A}$ is shown as the red curve in Figure~3. Also shown in blue is the conditional probability that an APT-accepted observation that achieves $m\ge 5$ valid Reference Stars, but does not achieve the safest `full house' allowance of $m=8$ 
$$p(k<8\vert k\ge5) = {{p(k\ge 5) - p(k\ge 8)}\over{p(k\ge 5)}} = {{p(k<8)-p(k<5)}\over{1-p(k<5)}} \eqno(5)$$
Figure~4 plots the same probabilities as a function of the density of Reference Star candidates on the sky for the June 2023 value of $p_{\scriptscriptstyle T\!A}=0.463$,

\topinsert
\vskip 0.0truecm
\centerline {\includegraphics[width=10.2cm, angle=0]{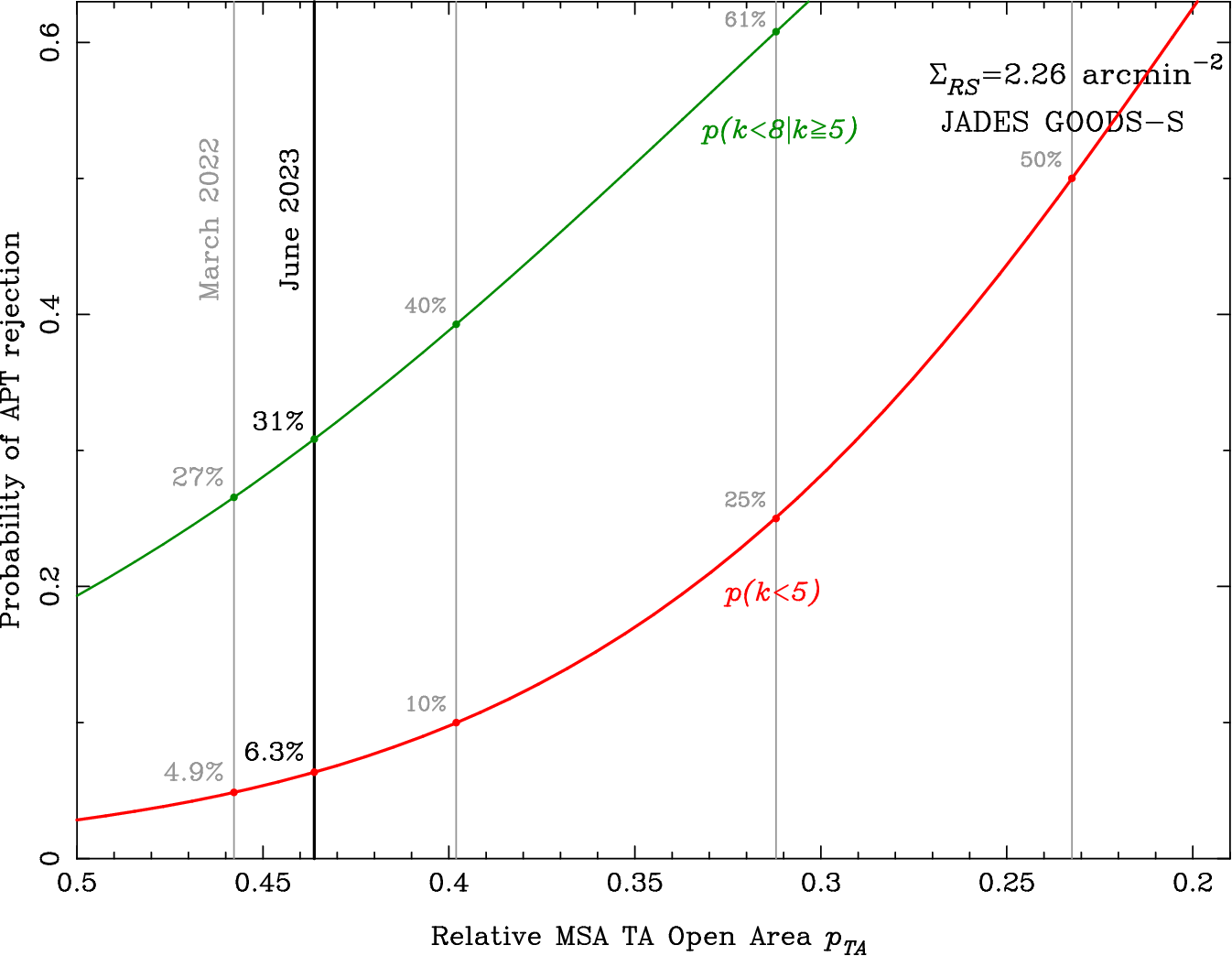}}
\vskip 0.5truecm
\centerline{\vbox{\hsize=13.5truecm {\bf Figure 3:} {\localcolor \Red Red} curve: Probability of not achieving the minimum number of $m=5$ Valid Reference Stars required by the APT as a function of the relative fraction of the MSA area available for target acquisition, $p_{\scriptscriptstyle T\!A}$, at a fixed candidate Reference Star surface density of $\Sigma_{\scriptscriptstyle R\!S}=2.26$~arcmin$^{-2}$. {\localcolor \Green Green} curve: Matching conditional probability of not achieving the maximum allowable number of $m=8$ valid Reference Stars given that at least $m=5$ are achieved. The values of $p_{\scriptscriptstyle T\!A}$ measured on March 2022 and June 2023 are indicated by vertical lines.}}
\endinsert

It is evident from Figure~3 that the probability that a given deep field NIRSpec MSA observation is rejected by the APT due to there being an insufficient number of Valid Reference Stars available grew from 4.9\% in March 2022 to 6.3\% in June 2023. Moreover, the proportion of accepted target acquisitions that do not achieve the maximum allowed number of Reference Stars increased from roughly one in four to nearly one in three over the same period. These numbers are in accord with actual experience on JADES (see however Section~6 below). This 28\% increase in the TA rejection rate represents an even stronger amplification of the base 1.9\% decrease in the number of functional shutters on the MSA.

\topinsert
\vskip 0.0truecm
\centerline {\includegraphics[width=10.2cm, angle=0]{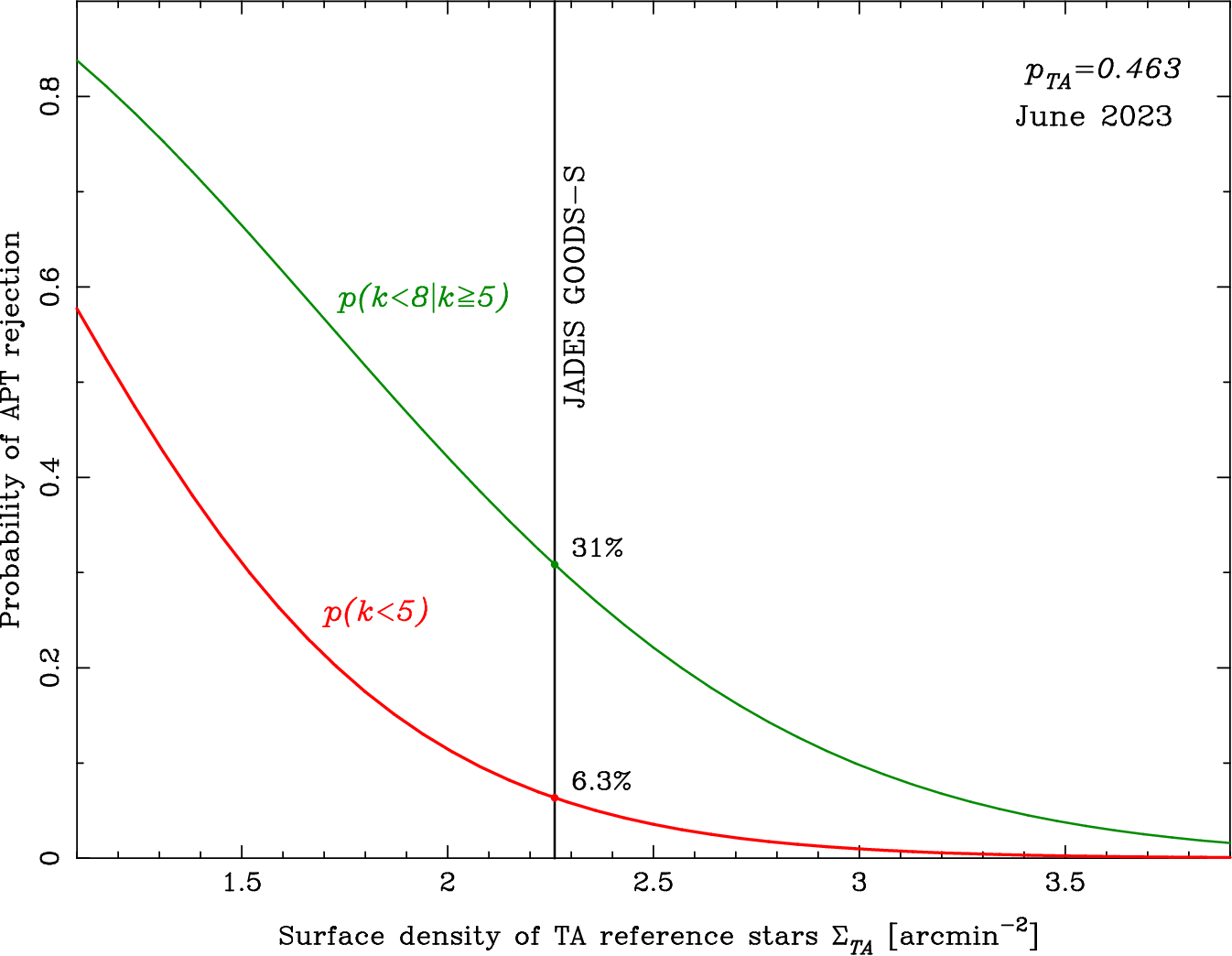}}
\vskip 0.5truecm
\centerline{\vbox{\hsize=13.5truecm {\bf Figure 4:} {\localcolor \Red Red} curve: Probability of not achieving the minimum number of $m=5$ Valid Reference Stars required by the APT as a function of the surface density of available Reference Stars candidates for a fixed value of $p_{\scriptscriptstyle T\!A}=0.463$. {\localcolor \Green Green} curve: Matching conditional probability of not achieving the maximum allowable number of $m=8$ valid Reference Stars given that at least $m=5$ are achieved. The nominal JADES value of $\Sigma_{\scriptscriptstyle R\!S}=2.26$~arcmin$^{-2}$ is indicated.}}
\vskip 0.3truecm
\endinsert

\smallskip
{\bf 6. Two potential relief valves}

While the APT rejecting a given planned MSA pointing due to its lack of Valid TA Reference Stars is a dire prospect, there are in truth two undocumented  potential work-arounds available, both of which have been successfully used on occasion on JADES. 

In many situations a given visit will be made up of several nearby dithered pointings that only require a single initial target acquisition. In such cases the APT assigns the target acquisition to the first pointing entered for the visit. However, as is clear from Figure~2, the pattern of Viable TA Windows changes on such small scales that it is entirely possible that a TA can be successfully executed at some of the close pointings, but not at others. The user can therefore try to re-enter the observation into the APT with the pointings in a different sequence in the hope that one will be accepted by the APT (however, the extreme tediousness of this trial-and-error process  has prompted the NIRSpec GTO team to develop a tool that pre-screens pointings for their Reference Star availability prior to APT entry).

The APT is remarkably unhelpful in informing the user of why it rejects or selects any given Reference Star candidate. If this occurs because the Reference Star does not fall in a Viable TA Window at the nominal pointing in question, then nothing can be done. However, if the Reference Star was rejected because its listed TA filter magnitude falls outside the allowed range for the assigned exposure time, it is possible to fool the APT into accepting the Reference Star by adjusting its claimed brightness in the user-supplied input catalog.

When the NIRSpec target acquisition approach was first proposed, and the first error budget for the scheme was constructed [5], it was pointed out that because of the very coarse 100~mas pixel sampling of the NIRSpec detector array, there is little point in exposing Reference Stars in the TA images to a detection level exceeding $S/N\simeq20$ since this will not result in an increase in the centroiding accuracy. However, during STScI's implementation of the scheme, $S/N=20$ was for some reason used to set the minimum brightness limits placed on the Reference Stars, when in reality the overall TA accuracy degrades gracefully for $S/N<20$.

Moreover, the Reference Star brightness limits given on JDOX are still the calculated pre-flight numbers, while the in-flight throughput of NIRSpec is known to be better than advertised before launch (an updated table did at one point briefly appear on JDOX, but was quickly replaced with the old table, and  was to our knowledge never implemented in the APT).

Therefore, while it is not advisable to fudge the listed TA filter magnitude of a Reference Star that is too bright according to the official range -- and therefore runs the risk of saturating on the detector -- it is in a pinch permissible to falsely brighten the listed entries of a limited number of Reference Stars that fall just below the allowed range by up to 0.5-1.0 magnitudes, with the aim of fooling the APT into accepting the observation in cases where a modest shortage of Valid Reference Stars is encountered. 

The above ploy has the effect of nudging the Reference Stars density slightly to the right in Figure~4. An increase in the number of available candidate Reference Stars could obviously also be obtained legally if STScI were to critically reassess its target acquisition magnitude limits and find grounds to revise them downward in brightness.

\bigskip
{\bf 7. Conclusion}

The 1.9\% loss of operational shutters on the MSA experienced on orbit between March 2022 and June 2023 has led to a modest 3.1\% reduction in the number of  high priority deep field science targets that NIRSpec is capable of observing simultaneously. This loss is not yet cause for concern. 

A greater worry is the impact on the ability of NIRSpec to perform autonomous MSATA target acquisition in deep fields, which is more sensitive to the number of randomly appearing failed closed shutters. In particular, the fraction of pointings that do not achieve the required minimum of five Valid Reference Stars within the MSA field of view has increased notably from 4.9\% to 6.3\%. At the same time, the percentage of higher risk target acquisitions that need to be carried out starting with fewer than the largest possible allocation of 8 Valid Reference Stars has increased from 27\% to 31\%. These numbers are not yet cause for panic, but do need to be carefully monitored going forward.

A naive linear extrapolation of the trend seen in the TA rejection probability suggests that 10\% of all deep field pointings will not be possible come December 2026. Whether such an interpolation is realistic is anyone's guess. It is encouraging that no new MSA short mask file has been issued since June 2023, which in part reflects the realization that some shorts appearing in the MSA are transient in nature. On the other hand, visual inspection of the most recent JADES MSA exposures taken in mid-January 2024 clearly show a number of shutter anomalies (including one new failed open shutter). This suggests that an update of the MSA operability tables may indeed be overdue.

\bigskip
{\bf Acknowledgement:} Tim Rawle is thanked for assistance in unravelling the intricacies of the MSA bookkeeping.

\bigskip
{\bf References}

[1] G. Giardino \& P. Ferruit, {\it Format definition - NIRSpec list of micro shutters operability} ESA-JWST-SCI-NRS-TN-2016-001. Issue 2.0, March 2016\hfill\break
\vskip -0.5truecm
[2] P. Ferruit  et al. 2022, A\&A, 661,A81, arXiv:2202.03306 \hfill\break
\vskip -0.5truecm
[3] C. D. Keyes et al. 2018, Proc. SPIE, 10704,107041J \hfill\break
\vskip -0.5truecm
[4] D. Eisenstein, et al. 2023 ApJS (submitted), arXiv:2306.02465\hfill\break
\vskip -0.5truecm
[5] P. Jakobsen, {\it Error Budget for NIRSpec Target Acquisition.} ESA-JWST-AN-3032, November 2005.\hfill\break

\bye